\documentclass[epj]{webofc}
\usepackage[varg]{txfonts}   
\usepackage{graphicx,color} 
\usepackage{bm}       
\usepackage{amsmath}  
\usepackage{amssymb}

\newcommand{\cut}[1]{}

\newcommand{\ket}[1]{\ensuremath{| #1 \rangle}}
\newcommand{\bra}[1]{\ensuremath{\langle #1  |}}

\woctitle{21st International Conference on Few-Body Problems in Physics}
\begin{document}
\title{Ab initio calculations of reactions with light nuclei}
\author{Sofia Quaglioni\inst{1}\fnsep\thanks{\email{quaglioni1@llnl.gov}} \and
        Guillaume Hupin\inst{2,1}\fnsep\thanks{\email{hupin@ipnorsay.in2p3.fr}} \and
        Angelo Calci\inst{3}\fnsep\thanks{\email{calci@triumf.ca}} \and
        Petr Navr\'atil\inst{3}\and Robert Roth\inst{4}
}

\institute{Lawrence Livermore National Laboratory, P.O.\ Box 808, L-414, 
Livermore, California 94551, USA 
\and
           Institut de Physique Nucl\'eaire, Universit\'ee Paris-Sud, IN2P3/CNRS, F-91406 Orsay Cedex, France 
\and
           TRIUMF, 4004 Wesbrook Mall, Vancouver, British Columbia, V6T 2A3, Canada
\and
	    Institut f\"ur Kernphysik, Technische Universit\"at Darmstadt, D-64289 Darmstadt, Germany	
          }

\abstract{
An {\em ab initio} (i.e., from first principles) theoretical framework capable of providing a unified description of the structure and low-energy reaction properties of light nuclei is desirable to further our understanding of the fundamental interactions among nucleons, and provide accurate predictions of crucial reaction rates for nuclear astrophysics, fusion-energy research, and other applications. In this contribution we review {\em ab initio} calculations for nucleon and deuterium scattering on light nuclei starting from chiral two- and three-body Hamiltonians, obtained within the framework of the {\em ab initio} no-core shell model with continuum. This is a unified approach to nuclear bound and scattering states, in which square-integrable energy eigenstates of the $A$-nucleon system are coupled to $(A-a)+a$ target-plus-projectile wave functions in the spirit of the resonating group method to obtain an efficient description of the many-body nuclear dynamics both at short and medium distances and at long ranges.} 
\maketitle
\section{Introduction}
\label{intro}
Understanding the structure and the dynamics of nuclei as many-body systems of protons and neutrons interacting through the strong force is one of the central goals of nuclear physics.  As systems that are amenable for accurate first-principle or {\em ab initio}  calculations, light nuclei represent an excellent testing ground in such an effort. Until about eight years ago, with the exception of studies involving $A=4$ nucleons or less, most of the progress in this direction has been achieved through {\em ab initio} calculations of well-bound nuclear states and their properties. However, efforts to generalize {\em ab initio} many-body techniques to also encompass $A>4$ nuclei that are weakly bound, or that only exist as low-energy resonances, as well as to explain light-nucleus scattering and low-energy reactions have been growing, driven by new ideas, the rising availability of data on exotic nuclei, and increased access to high-performance computing. At the same time, effective field theory (EFT) and renormalization group transformations are providing nuclear theorists with a systematic approach to, and a much-improved understanding of the nuclear forces, and are yielding increasingly precise interactions and theoretical uncertainties, which are the basis for accurate predictions. 
The combination of these developments is opening 
the exciting prospect of a predictive and quantified theory of light-nucleus reactions important for applications, e.g.:  scattering processes widely used in material science to probe thin films, or analyze the surface of artworks and archeological artifacts; thermonuclear fusion reactions crucial for big-bang and stellar nucleosyntesis; or those that power research facilities directed toward developing fusion energy. 
For the success of such a program it is essential to properly understand three key elements and their interplay: the role of nucleon-nucleon (NN) and three-nucleon (3N) forces; the influence of open channels and continuum degrees of freedom; and the propagation of the uncertainties in interaction models into error bars on computed observables. In this contribution we will review recent progress in addressing the first two of these elements by means of {\em ab initio} calculations of light-nucleus structure and dynamics carried out within the no-core shell model with continuum (NCSMC) approach. In particular, in Sec.~\ref{sec-t} we will briefly outline the main aspects of the NCSMC formalism, while results for nucleon- and deuterium-nucleus dynamics employing chiral EFT NN+3N forces will be discussed in Sec.~\ref{sec-r}. Concluding remarks and future prospects will be presented in Sec.~\ref{sec-c}.

\section{Formalism}
\label{sec-t}
To achieve a unified {\em ab initio} description of structural and reaction properties of light nuclei, we work within the theoretical framework of the NCSMC, introduced in Refs.~\cite{Baroni2013} and \cite{Baroni2013a}.   
In this approach, the translational-invariant ansatz for the many-body wave function in each partial wave of total angular momentum $J$, parity $\pi$ and isospin $T$,  is given by a superposition of square-integrable energy eigenstates of the compound $A$-nucleon system and continuous cluster states according to: 
\begin{align}
\label{NCSMC_wav}
\ket{\Psi^{J^\pi T}_A} & =  \sum_\alpha c^{J^\pi T}_\alpha \ket{A \alpha J^\pi T} \! +\! \sum_{\nu}\! \int \! dr \: r^2 
                               \frac{\gamma_{\nu}^{J^\pi T}(r)}{r}
                               {\mathcal{A}}_\nu\ket{\Phi_{\nu r}^{J^\pi T}}\,.
\end{align}
Here $\ket{A \alpha J^\pi T} $ represent antisymmetric, translation-invariant discrete eigenstates of the $A$-nucleon Hamiltonian, labelled by the energy index $\alpha$. They are obtained ahead of time by means of the {\em ab initio} no-core shell model (NCSM)~\cite{Navratil2000a} through the diagonalization of the $A$-nucleon microscopic Hamiltonian within a model space spanned by many-body harmonic oscillator (HO) wave functions with up to $N_{\rm max}$ HO excitations above the unperturbed configuration and frequency $\hbar\Omega$. Further  
\begin{align}
\label{RGMbasis}
	\ket{\Phi_{\nu r}^{J^\pi T}} = &
	\Big[ \left(
         \ket{A-a \; \alpha_1 I_1^{\pi_1}T_1}\ket{a \; \alpha_2 I_2^{\pi_2}T_2}
         \right)^{(sT)} 
         Y_\ell(\hat{r}_{A-a,a})
         \Big]^{(J^{\pi}T)}  \frac{\delta(r-r_{A-a,a})}{rr_{A-a,a}} \; .
\end{align}
are continuous 
resonating-group method (RGM)~\cite{Tang1978, Langanke1987} basis states, representing an $(A-a)$-nucleon target and $a$-nucleon projectile (with $a \le A$), whose centers of mass are separated by the relative coordinate $\vec r_{A-a,a}$, traveling in a $^{2s+1}\ell_J$ wave of relative motion (with $s$ the channel spin, and $\ell$ the relative momentum of the system). The eigenstates $ \ket{A-a \; \alpha_1 I_1^{\pi_1}T_1}$ and $\ket{a \; \alpha_2 I_2^{\pi_2}T_2}$ of each cluster of nucleons are obtained analogously to, and consistently with those of the $A$-nucleon system and are characterized by total angular momentum $I_{1,2}$, parity $\pi_{1,2}$, isospin $T_{1,2}$, and energy index $\alpha_{1,2}$. The index $\nu=\{A{-}a\,\alpha_1I_1^{\,\pi_1} T_1;\, a\, \alpha_2 I_2^{\,\pi_2} T_2;\, s\ell\}$ collects all quantum numbers associated with this continuous basis, and its full antisymmetrization 
is recovered by introducing an appropriate inter-cluster antisymmetrizer ${\mathcal{A}}_\nu$, which accounts for exchanges of nucleons between target and projectile. The unknown discrete $c^{J^\pi T}_\alpha$ and continuous $\gamma_{\nu}^{J^\pi T}(r) = ({\mathcal N}^{-1/2}\chi)_\nu(r)$ variational amplitudes, are the solutions of the orthogonalized coupled-channel equations 
\begin{eqnarray}
\left(
\!\!\!
\begin{array}{cc}
        E_{\alpha}\,\delta_{\alpha\alpha^\prime} & (h{\mathcal N}^{-\frac12})_{\alpha \nu^\prime}(r^\prime)
 	\\[4mm]  
	(h{\mathcal N}^{-\frac12})_{\alpha^\prime \nu}(r)
        & (\mathcal{N}^{-\frac12}\mathcal{H}\mathcal{N}^{-\frac12})_{\nu\nu^\prime}(r,r^\prime)
\end{array}
\!\!\right) \left(
\!\!\!\begin{array}{c}
	c_{\alpha^\prime} \\[4mm]
	 \frac{\chi_{\nu^\prime}(r^\prime)}{r^\prime}
\end{array}
\!\!\!\right)  =   E \left(
\!\!\!\begin{array}{cc}
        \delta_{\alpha\alpha^\prime} \!\!&\!\! (g{\mathcal N}^{-\frac12})_{\alpha \nu^\prime}(r^\prime) 
        \\[4mm]  
        (g{\mathcal N}^{-\frac12})_{\alpha^\prime \nu}(r)
        \!\!&\!\! \frac{\delta_{\nu\nu^\prime}\delta(r-r^\prime)}{r r^\prime}
\end{array}
\!\!\!\right) \left(
\!\!\!\begin{array}{c}
c_{\alpha^\prime} \\[4mm]
 \frac{\chi_{\nu^\prime}(r^\prime)}{r^\prime}
\end{array}
\!\!\!
\right)\,,
\label{eq:NCSMC-eq}
\end{eqnarray}
which follow from the many-body Schr\"odinger equation when using the ansatz~(\ref{NCSMC_wav}) for the many-body wave function and projecting over the model space spanned by the  $\ket{A \alpha J^\pi T} $ and \ket{\Phi_{\nu r}^{J^\pi T}} basis states.
In Eq.~(\ref{eq:NCSMC-eq}), $E$ denotes the total energy of the system and the two by two block-matrices on the left- and right-hand side of the equation represent, respectively, the NCSMC Hamiltonian and norm (or overlap) 
kernels.  The upper diagonal blocks are given by the Hamiltonian (overlap) matrix elements over the square-integrable part of the basis. In particular, as the basis states are NCSM eigenstates of the $A$-nucleon Hamiltonian, these are trivially given by the diagonal matrix of the $E_\alpha$ eigenergies (the identity matrix). Similarly, those over the orthonormalized continuous portion of the basis appear in the lower diagonal block and are obtained from the norm (or overlap) RGM kernel 
${\mathcal N}_{\nu\nu^\prime}(r,r^\prime)\!=\!\bra{\Phi^{J^\pi T}_{\nu r}} {\mathcal A}_\nu{\mathcal A}_{\nu^\prime}\ket{\Phi^{J^\pi T}_{\nu^\prime r^\prime}}$ and Hamiltonian kernel ${\mathcal H}_{\nu\nu^\prime}(r,r^\prime)\!=\!\bra{\Phi^{J^\pi T}_{\nu r}} {\mathcal A}_\nu H {\mathcal A}_{\nu^\prime}\ket{\Phi^{J^\pi T}_{\nu^\prime r^\prime}}$~\cite{Quaglioni2008,Quaglioni2009}.
The off diagonal blocks contain the couplings between the two sectors of the basis, with $g_{\alpha \nu}(r)\!=\!\bra{A\, \alpha J^\pi T} {\mathcal A}_{\nu}\ket{\Phi^{J^\pi T}_{\nu r}}$ and $h_{\alpha\nu}(r)\!=\!\bra{A \, \lambda J^\pi T}H {\mathcal A}_{\nu}\ket{\Phi^{J^\pi T}_{\nu r}}$. 
The scattering matrix (and from it any scattering observable) follows from matching the solutions of Eq.~(\ref{eq:NCSMC-eq}) with the known asymptotic behavior of the wave function at large distances by means of the microscopic $R$-matrix method~\cite{Hesse1998,Hesse2002}.

\section{Unified description of structure and reaction observables with chiral two- and three-nucleon forces}
\label{sec-r} 
The {\em ab initio} NCSMC was initially developed to compute nucleon-nucleus collisions starting from a two-body Hamiltonian and applied to the description of the unbound $^7$He nucleus~\cite{Baroni2013,Baroni2013a}. The approach was  later extended to include explicit 3N forces, which are necessary to obtain a truly accurate and quantitative description of light-nucleus scattering process~\cite{Nollett2007,Viviani2013}. In the following we review applications of such an extended approach, using an Hamiltonian based on the chiral N$^3$LO  NN interaction of Ref.~\cite{Entem2003} and N$^2$LO  3N force of Ref.~\cite{Navratil2007}, constrained to provide an accurate description of the $A=2$ and $3$~\cite{Gazit2009} systems and unitarily softened via the similarity-renormalization-group (SRG) method~\cite{Glazek1993,Wegner1994,Bogner2007,Hergert2007,Jurgenson2009} to minimize the influence of momenta higher than a low-momentum resolution scale of $\Lambda_{\rm SRG}=2.0$ fm$^{-1}$. In particular, Sec.~\ref{nucleon} discusses predictions of elastic scattering and recoil of protons off $^4$He~\cite{Hupin2014}, as well as a study of continuum and $3N$-force effects on the energy levels of $^9$Be~\cite{Langhammer2015}. In Sec.~\ref{deuteron} we review the first application to describe more challenging deuterium-nucleus dynamics~\cite{Hupin2015}, and discuss our ongoing effort to describe deuteron-induced nucleon transfer reactions.

\subsection{Nucleon-nucleus scattering}
\label{nucleon}

The recent NCSMC study of low-energy cross sections for elastic scattering and recoil of protons from $^4$He of Ref.~\cite{Hupin2014} is an excellent example of how the combination of efficient {\em ab initio} approaches, accurate nuclear interactions and high-performance computing capabilities can lead to a predictive theory of light-nucleus reactions important for applications. In this case, accurate $^4$He$(p,p)^4$He and $^1$H$(\alpha,p)^4$He angular differential cross sections for a variety of proton/$^4$He incident energies and detection angles are key to the feasibility and quality of ion-beam analysis applications aimed at determining the concentrations and depth profiles of helium and hydrogen, respectively, at the surface of materials or in thin films. Starting from SRG-evolved chiral NN+3N forces with $\Lambda_{\rm SRG}=2.0$ fm$^{-1}$, accurate results for $p-^4$He scattering were obtained by employing an $N_{\rm max}=13$, $\hbar\Omega=20$ MeV HO model space, proton-$^4$He binary-cluster basis states including up to the first seven ($I_1^{\pi_1} T_1 =$ 0$_1^{\texttt+}$0, 0$_2^{\texttt+}$0, 0$^{\texttt-}$0, 2$^{\texttt-}$0, 2$^{\texttt-}$1, 1$^{\texttt-}$1 and 1$^{\texttt-}$0) $^4$He eigenstates, and the first fourteen (of which three $3/2^{\texttt-}$ and two $1/2^{\texttt-}$) square-integrable eigenstates of the $^5$Li compound nucleus.
\begin{figure}[t]
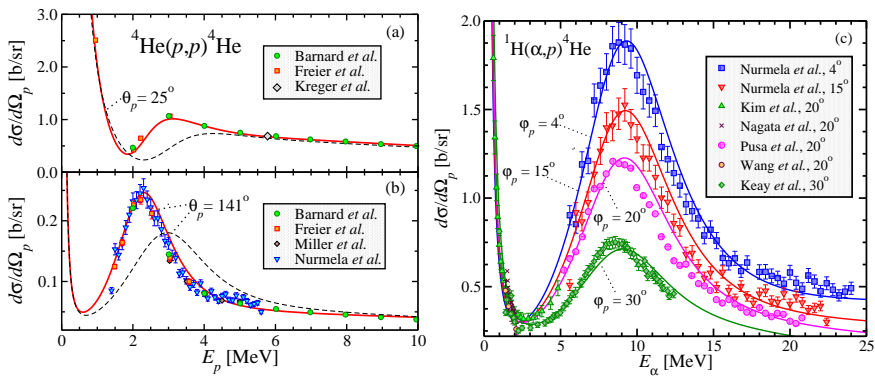

\begin{center}
\includegraphics[width=.39\textwidth,clip]{QuaglioniS_fig1ab.eps}
\includegraphics[width=.41\textwidth,clip]{QuaglioniS_fig1c.eps}
\end{center}
\caption{Computed 
(lines) $^4$He$(p,p)^4$H angular differential cross section at forward scattering angle $\theta_p=25^\circ$ (a) and backscattering angle $\theta_p=141^\circ$ (b) as a function of the proton incident energy compared with measurements (symbols) from Refs.~\cite{Freier1949,Kreger1954,Miller1958,Barnard1964,Nurmela1997}, and $^1$H$(\alpha,p)^4$He angular differential cross section at the proton recoil angles $\varphi_p=4^\circ, 16^\circ, 20^\circ$, and $30^\circ $($\rm{c}$) as a function of the incident $^4$He energy compared with data (symbols) from Refs.~\cite{Nagata1985,Baglin1992,Wang1986,Nurmela1997,BogdanovicRadovic2001,Kim2001,Keay2003,Pusa2004,Browning2004}. The 
solid lines corresponds to the present NCSMC calculations. Also shown (black dashed lines) are the results of Ref.~\cite{Hupin2013}, i.e.\ without $^5$Li square intregrable eigenstates.}
\label{figure1}
\end{figure}
As can be seen from Fig.~\ref{figure1}
the computed NCSMC angular differential cross sections agree very well with data both for $^4$He$(p,p)^4$He and $^1$H$(\alpha,p)^4$He processes. For comparison, in panels (a) and (b) we also show as black dashed lines the results~\cite{Hupin2013} obtained within a model space spanned by $p-^4$He cluster states alone, that is by retaining only the second term of Eq.~(\ref{NCSMC_wav}) and solving $\mathcal{N}^{-\frac12}\mathcal{H}\mathcal{N}^{-\frac12} \, \chi=E\,\chi$. While the high-energy tail of the cross section can already be described well working within such more limited model space, the inclusion of square-integrable eigenstates of the compound $^5$Li system is essential at lower energy, where it efficiently compensates for missing higher excitations of the $^4$He core.

\begin{figure}[t]
\centering
\sidecaption
\includegraphics[width=0.4\textwidth,clip]{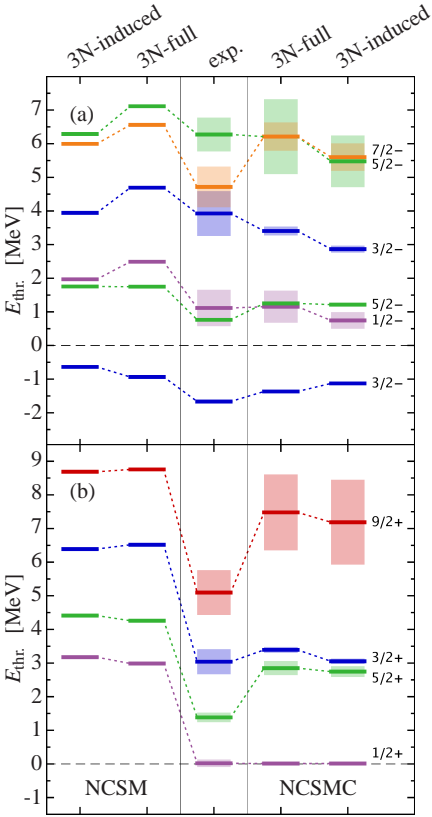}
\caption{Computed low-lying energy spectrum of $^9$Be (a) negative- and (b) positive-parity states compared to experiment. The chiral $NN+3N$ Hamiltonian of Ref.~\cite{Roth2012} with evolution parameter $\lambda= 2$ fm$^{-1}$ was used. The negative(positive)-parity NCSMC calculation, coupling $^9$Be square integrable eigenstates with $^8$Be$(0^+,2^+)$ binary cluster states, was performed in a $N_{\rm max}=12(11)$, $\hbar\Omega=20$ MeV HO model space. Here $E_{\rm kin}$ indicates the center-of-mass energy with respect to the $^8$Be+$n$ threshold.}
\label{figure2}
\end{figure}

At the same time, a proper treatment of continuum degrees of freedom is also indispensable to draw reliable conclusions on the influence of 3N forces in the low-lying spectra of loosely bound systems, such as $^9$Be, for which all excited states lie above the $n$+$^8$Be threshold. The positive parity resonances of this nucleus are in general found too high in energy compared to experiment, in {\em ab initio} calculations that treat them as bound states such in the NCSM study of Ref.~\cite{Forssen2005}. In the same setting, counter to expectations, the splitting between the lowest $5/2^-$ and $1/2^-$ resonances tends to be overestimated when $3N$ effects are included. A recent NCSMC study of $^9$Be as a linear combination of 9-body square-integrable eigenstates and $^8$Be+$n$ binary-cluster states with the $^8$Be in its ground and $2^+$ states~\cite{Langhammer2015} helped shed some light on the interplay of continuum degrees of freedom and 3N force effects in this nucleus. As can be seen in Fig.~\ref{figure2}, for all energy levels the inclusion of the continuum significantly improves the agreement with experiment. In particular, the splitting between the $5/2^-$ and $1/2^-$ levels is substantially reduced when the continuum is included due to a shift towards lower energies of the $^2P_{1/2}$ resonance. However, the most dramatic continuum effects were found in the positive-parity resonances, shown in Fig.~\ref{figure2}(b). The $1/2^+$ and the $3/2^+$ $S$-wave resonances are several MeV lower in the NCSMC, close to their experimental value.


\subsection{Deuterium-induced reactions}
\label{deuteron}
One of the most challenging and notable applications of the NCSMC accomplished so far is the simultaneous study of the $^6$Li ground state and $d-^4$He elastic scattering using NN+3N forces from chiral effective field theory (softened -- as in the previous section -- by means of an SRG transformation with $\Lambda_{\rm SRG}=2.0$ fm$^-1$)~\cite{Hupin2015}. Unless otherwise stated, all calculations were carried out using the ansatz of Eq.~(\ref{NCSMC_wav}) with fifteen discrete eigenstates of the $^6$Li system and continuous $d$-$^4$He(g.s.)\ binary-cluster states with up to seven deuteron pseudostates in the $^3S_1{-}{}^3D_1$, $^3D_2$ and $^3D_3{-}{}^3G_3$ channels. Convergence in the HO expansions was approached at at $N_{\rm max}=11$, using a frequency of $20$ MeV,  
around which the $^6$Li g.s.\ energy calculated within the square-integrable basis of the NCSM becomes nearly insensitive to $\hbar\Omega$~\cite{Jurgenson2011}. 
 
\begin{figure}[t]
\begin{center}
\includegraphics[width=.5\textwidth,clip]{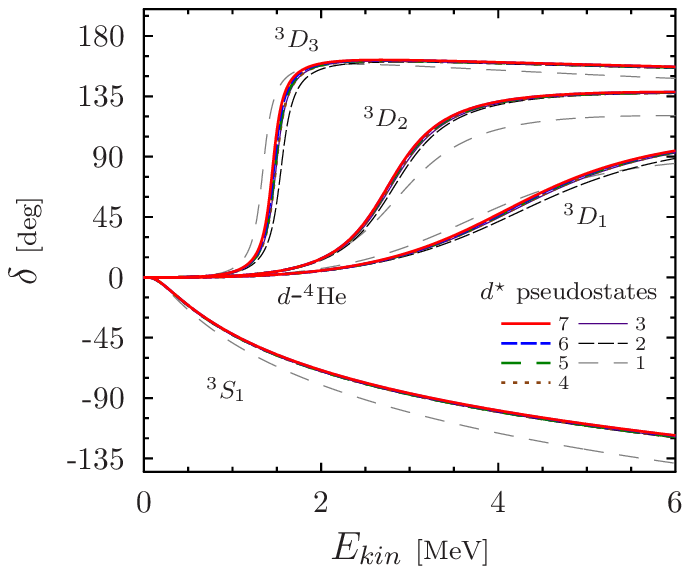}
\includegraphics[width=.42\textwidth,clip]{QuaglioniS_fig3b.eps}
\end{center}
\caption{Panel (a): Computed $d$-$^4$He $S$- and $D$-wave phase shifts at $N_{\rm max}=9$, 
obtained  with fifteen square-integrable $^6$Li eigenstates, as a function of the number of $^2$H pseudostates (up to seven) in each of the $^3S_1{-}{}^3D_1$, $^3D_2$ and $^3D_3{-}{}^3G_3$ channels. The two-body part of the SRG-evolved N$^3$LO $NN$ potential ($NN$-only) with $\Lambda=2.0$ fm$^{-1}$ was used.
Panel (b): Computed $^2$H$(\alpha,d)^4$He using the $NN+3N$ Hamiltonian (lines) and measured (symbols) center-of-mass angular distributions at $E_d=2.93,6.96, 8.97$~\cite{Jett1971}, and $12$ MeV~\cite{Senhouse1964}, scaled by a factor of $20,5,2$, and $1$, respectively. All positive- and negative-parity partial waves up to $J=3$ were included in the calculations.}
\label{figure3}
\end{figure}

Similar to the proton-$^4$He scattering discussed earlier, also in this case the square-integrable eigenstates of the compound (here $^6$Li) system play a crucial role in achieving an accurate description of the dynamic process.  Besides helping to account for the polarization of the $^4$He, of which we can (computationally) afford to include only the g.s., here they also contribute to the description of the projectile distortion. This is exemplified by the fairly rapid convergence, shown in the left panel of Fig.~\ref{figure3}, of the $d-^4$He scattering phase shifts with respect to the number of deuterium pseudostates. The high quality of the results obtained with the chiral NN+3N forces is corroborated by the good agreement with experiment of Fig.~\ref{figure3} (right panel), presenting $^4$He$(d,d)^4$He angular distributions in the $2.93\le E_d\le 12.0$ MeV interval of incident energies.

\begin{figure*}[t]
\begin{minipage}{0.49\textwidth}
\begin{center}
\includegraphics[width=1\textwidth]{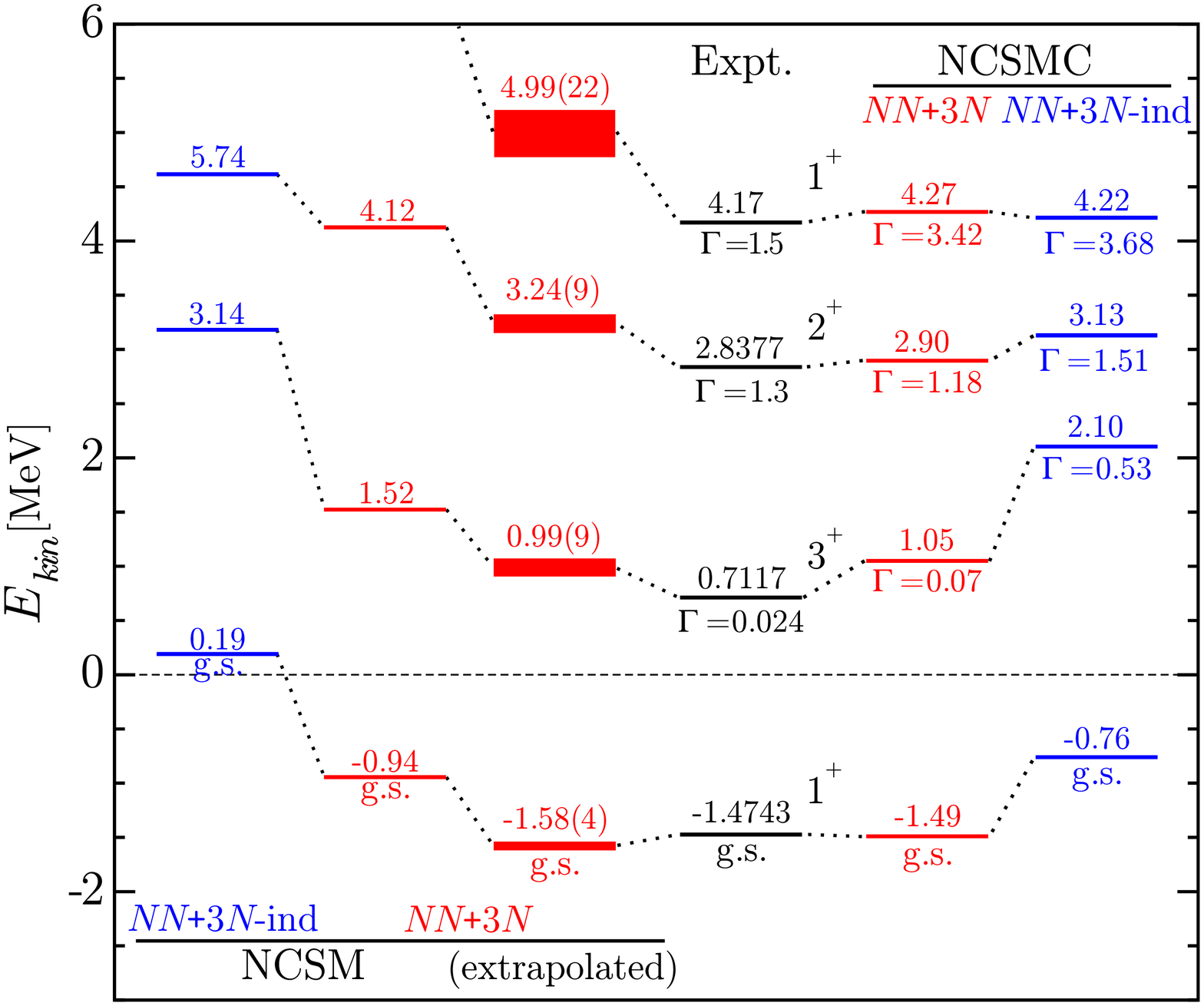}
\caption{Computed NCSMC low-lying spectrum of positive-parity states (and their widths $\Gamma$) for the $^6$Li nucleus with (red lines) and without (blue lines) inclusion of the initial chiral $3N$ force compared to experiment (black lines). Also shown on the left-hand side are the best ($N_{\rm max}=12$) and extrapolated ($N_{\rm max}=\infty$) NCSM energy levels. The zero energy is set to the respective computed (experimental) $d+^4$He breakup thresholds.}
\label{figure4}
\end {center}
\end{minipage}\hspace{2pc}%
\begin{minipage}{0.49\textwidth}
\begin{center}
\includegraphics[width=1\textwidth]{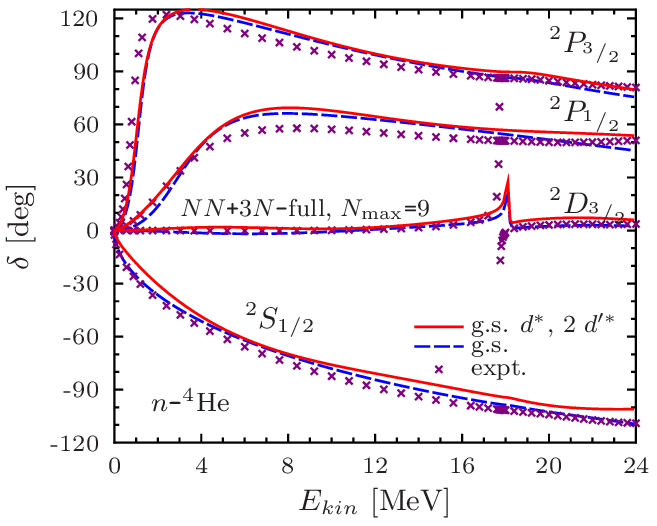}
\caption{Preliminary results (lines) for the $n-^4$He scattering phase shifts from zero to 24 MeV in the center-of-mass energy compared to experiment (crosses). All calculations were performed at $N_{\rm max}=9$ within an over-complete NCSMC model space including $n+^4$He(g.s.)\ and $d+^3$H(g.s.)\ continuous basis states with up to two $^2$H pseudostates in the  $^3S_1$--$^3D_1$  $(d^*)$ and $^3D_2$  $(d^{\prime*})$ channels, as well as square-integrable discrete eigenstates of the compound $^5$He nucleus.}
\label{figure5}
\end{center}
\end{minipage} 
\end{figure*}

At the same time, the inclusion of continuum degrees of freedom is essential to reach a true understanding of a nucleus with low breakup threshold such as $^6$Li, the ground state of which lies only 1.47 MeV (compared to its absolute energy of nearly 32 MeV) below the $^4$He$+d$ separation energy. As shown in Fig.~\ref{figure4}, combined with the inclusion of the $3N$ force, the NCSMC yields a rather good agreement with the observed spectrum. As one should expect, for the ground state and the narrow $3^+$ resonance the NCSMC energy levels are also in good agreement with those obtained from an extrapolation to $N_{\rm max}=\infty$ of a traditional NCSM calculation. This is a strong indication that the slight overestimation of the first excited state is likely due to remaining deficiencies in the adopted 3N force model, particularly concerning the strength of the spin-orbit interaction. However, only within the NCSMC do the computed wave functions present the correct asymptotic, which for the g.s. are Whittaker functions. This is essential for the extraction of the asymptotic normalization constants. The present calculation reproduces the empirical binding energy of $^6$Li, yielding an asymptotic D- to S-state ratio of the $^6$Li wave function in $d+\alpha$ configuration of $-0.027$ in agreement with a determination from $^6$Li-$^4$He elastic scattering~\cite{George1999}. Contrary to the lighter nuclei, this ratio was still uncertain for $^6$Li, with different determinations disagreeing even as to its sign~\cite{Tilley2002a}.

The above accurate studies of nucleon- and deuterium-$^4$He elastic scattering with chiral NN+3N interactions set the stage for the most advanced {\em ab initio} calculation of the $^3$H$(d,n)^4$He fusion, currently under way. This study is being carried out within an over-complete NCSMC model space including $n+^4$He(g.s.) and $d+^3$H(g.s.) continuous basis states, as well as square-integrable discrete eigenstates of the compound $^5$He nucleus. Figure~\ref{figure5} shows preliminary results for the $n-^4$He scattering phase shifts from zero to 24 MeV in the center-of-mass energy. Despite the fairly small size of the harmonic oscillator basis, the calculation is in close agreement with experiment. In particular, besides a slight shift of the $P$-wave resonances to lower energies, the inclusion of $d+^3$H channels leads to the appearance of a resonance in the $^2D_{3/2}$ partial wave, just above the $d+^3$H threshold. This is the exit channel of the deuterium-tritium fusion. 

\section{Conclusions and future prospects}
\label{sec-c}
A unified {\em ab initio} description of light-nucleus structure and reaction properties starting from accurate NN+3N forces is now becoming possible, thanks in part to the NCSMC approach. This is an {\em ab initio} theory including the continuum which combines the efficient description of short- and medium-range correlations of the NCSM with the ability of the RGM of describing the scattering physics of a system. In this contribution, we reviewed recent applications of the approach to the description of elastic scattering and recoil of protons from $^4$He, continuum and 3N-force effects on the energy levels of the $^9$Be, and $^6$Li structure and $d-^4$He dynamics using chiral NN+3N forces. Building on this work, we are now pursuing several new applications to light-nucleus structure and reaction properties with chiral NN+3N forces. The obtained accurate nucleon-$^4$He scattering wave functions are currently being used as input for the study of the $^4$He$(p,p^\prime \gamma)^4$He and $^4$He$(n,n^\prime \gamma)^4$He bremsstrahlung, the radiative process by which a photon is emitted as a result of the nuclear collision between a nucleon and a $^4$He nucleus. The former is one of the few measured light-nucleus bremsstrahlung cross sections, while the latter is a necessary preliminary step for the study of the more complicated $^3$H$(d,n\gamma)^4$He bremsstrahlung cross section, which could be used to diagnose plasmas in fusion experiments but is not known well enough. The general NCSMC framework applicable to targets heavier than $^4$He is now being applied to study continuum and 3N force effects on the low-lying spectrum of the $^{11}$Be one-neutron halo nucleus and its photodissociation into $n+^{10}$Be. The  obtained accurate $^6$Li ground state and $d-^4$He scattering wave functions set the stage for the first {\em ab initio} study of the $^2$H$(\alpha,\gamma)^6$Li radiative capture, responsible for the big-bang nucleosynthesis of $^6$Li, as well as the calculation of $^6$Li g.s.\ properties including the effect of the continuum degrees of freedom. Finally, we are in the process of  carrying out the first {\em ab initio} calculation of the $^3$H$(d,n)^4$He fusion using chiral NN+3N forces.

\begin{acknowledgement}
Prepared in part by LLNL under Contract DE-AC52-07NA27344. This material is based upon work supported by the U.S.\ Department of Energy, Office of Science, Office of Nuclear Physics, under 
Work Proposal Number SCW1158,  
the NSERC Grant Number 401945-2011, the
Deutsche Forschungsgemeinschaft through Contract SFB 634,
the Helmholtz International Center for FAIR within the
framework of the LOEWE program launched by the State of
Hesse, and the BMBF through Contract No. 06DA7074I. TRIUMF receives funding via a contribution through the 
Canadian National Research Council.
Computing support for this work 
came from the LLNL institutional Computing Grand Challenge
program, the National Energy Research Scientific Computing Center (edison) supported by the Office of Science of the U.S. Department of Energy under Contract No. DE-AC02-05CH11231, the LOEWE-CSC Frankfurt, the computing center of the TU Darmstadt (lichtenberg), and from an INCITE Award on the Titan supercomputer of the Oak Ridge Leadership Computing Facility (OLCF) at ORNL.
\end{acknowledgement}

%

\end{document}